\documentclass[prl,aps,twocolumn,showpacs]{revtex4}

\usepackage{graphicx}
\usepackage{bm}

\begin{document}

\title{Real-time dynamics in Quantum Impurity Systems: A
Time-dependent Numerical Renormalization Group Approach}

\smallskip

\author{Frithjof B.~Anders$^1$  and Avraham Schiller$^2$}
\affiliation{$^1$Department of Physics, Universit\"at Bremen,
                  P.O. Box 330 440, D-28334 Bremen, Germany\\
             $^2$Racah Institute of Physics, The Hebrew University,
                  Jerusalem 91904, Israel}
\date{May 19, 2005}

\begin{abstract}
We develop a general approach to the nonequilibrium dynamics of
quantum impurity systems for arbitrary coupling strength. The
numerical renormalization group is used to generate a complete
basis set necessary for the correct description of the time
evolution. We benchmark our method with the exact analytical
solution for the resonant-level model. As a first application,
we investigate the equilibration of a quantum dot subject to a
sudden change of the gate voltage and external magnetic field.
Two distinct relaxation times are identified for the spin and
charge dynamics.
\end{abstract}

\maketitle


\paragraph{Introduction.}
The investigation of real-time dynamics in quantum impurity
systems (QIS) is of prime importance for our understanding
of dissipation and decoherence in qubits and electronic
transport through nano-devices. Such systems consist of
a mesoscopic subsystem or device, interacting with an
infinitely large environment of noninteracting particles.
Typical examples are quantum dots, qubits or biological
donor-acceptor molecules. These are usually modeled either
by a spin-Bose model, where a two-level system or qubit is
interacting with a Bosonic bath~\cite{Leggett1987}, or
by a set of atomic orbitals coupled to electronic
leads, as in ultra-small quantum dots and single-electron
transistors~\cite{KastnerSET1992}.
Recently, Elzerman {\em et al.}~\cite{Kouwenhoven2004}
reported the usage of gate-voltage pulses for a single-shot
read out of the spin configuration of a single-electron
transistor in a finite magnetic field. Such devices may
be suitable for quantum computing, as they combine a
long-lived quantum state with an easy read-out protocol
furnished by the coupling to the environment.


In contrast to equilibrium conditions, the understanding of
the real-time evolution of many-particle quantum systems is
still at its infancy. Over the past 40 years, the Keldysh
technique~\cite{Keldysh65} has proven to be the most
successful approach to nonequilibrium (NEQ) dynamics, as it
furbishes a perturbative expansion of the density operator.
In general, however, perturbation theory fails in QIS due
to the infra-red divergences caused by degeneracies on the
impurity. Recently, there has been significant progress in
extending the density matrix renormalization group (DMRG)
to time-dependent 1D quantum systems~\cite{Marston2002,
DaleyKollathSchollwoeckVidal2004, WhiteEeiguin2004,
GobertTdDMRG2004}.
The so-called TD-DMRG works well for finite-size systems
and short times, but has an accumulated error proportional 
to the time elapsed. This makes the TD-DMRG unsuitable at
present for tackling long time scales, most notably the
exponentially long time associated with the development
of the Kondo effect~\cite{NordlanderEtAl1999}.


In this Letter, we develop an alternative approach to
the real-time dynamics of QIS based on Wilson's numerical
renormalization-group (NRG) method~\cite{Wilson75}. The
NRG is a very powerful and accurate tool for calculating
equilibrium properties of arbitrarily complex quantum
impurities. We extend the approach to a certain
class of time-dependent problems where a sudden
perturbation, e.g., a gate-voltage
pulse~\cite{Kouwenhoven2004}, is applied to the impurity
at time $t = 0$. An earlier attempt~\cite{Costi97} at a
time-dependent NRG (TD-NRG) was faced with conceptual
problems how to adequately couple the low- and high-energy
scales. Here we resolve these problems by combining the
NRG with Feynman's concept of a reduced density
matrix~\cite{Feynman72,Hofstetter2000}, and by implementing
a suitable resummation procedure for tracking all states
that contribute to the time evolution of the system. We
establish the accuracy of our approach by comparison to
the exact analytical solution of the resonant-level model,
and apply it to the single impurity Anderson model (SIAM).
Two distinct time scales are found for the spin and charge
dynamics in the SIAM.
The process of spin relaxation is shown to be sensitive
to the initial conditions imposed on the system.


\paragraph{Formulation.}
As our approach applies to arbitrary QIS, we first
formulate it in general terms before turning to concrete
examples. The Hamiltonian of QIS is generally given
by ${\cal H}^i = {\cal H}_{bath} + {\cal H}_{imp} +
{\cal H}_{mix}$, where ${\cal H}_{bath}$ models the
continuous bath, ${\cal H}_{imp}$ represents the decoupled
impurity, and ${\cal H}_{mix}$ describes the coupling
between the two subsystems. The entire system is assumed
to be in thermal equilibrium for $t < 0$, at which point
a sudden perturbation $\Delta {\cal H}$ is switched on:
${\cal H}(t>0) = {\cal H}^i + \Delta {\cal H} \equiv
{\cal H}^f$. For $t > 0$, the density-matrix operator evolves
according to $\hat{\rho}(t > 0) = e^{-it {\cal H}^f}
\hat{\rho}_{eq} e^{it {\cal H}^f}$, where $\hat{\rho}_{eq}
= e^{-\beta {\cal H}^i}/ {\rm Tr}\{e^{-\beta {\cal H}^i}\}$.

The key ingredient in the NRG is a logarithmic discretization
of the continuous bath, controlled by the parameter
$\Lambda > 1$~\cite{Wilson75}. The Hamiltonian is mapped
onto a semi-infinite chain, where the $N$th link
represents an exponentially decreasing energy scale
$D_N \sim \Lambda^{-N/2}$~\cite{Wilson75}. Using this
hierarchy of scales the sequence of finite-size
Hamiltonians ${\cal H}_N$ is solved iteratively, truncating
the high-energy states at the conclusion of each step to
maintain a manageable number of states. The reduced basis
set of ${\cal H}_N$ so obtained is expected to faithfully
describe the spectrum of the full Hamiltonian on a
scale of $D_N$, corresponding to the temperature
$T_N \sim D_N$~\cite{Wilson75}.

\begin{figure}[btp]
\centering
\includegraphics[width=80mm]{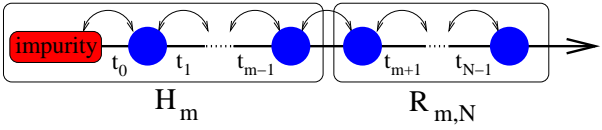}
\caption{The full NRG chain of length $N$ is divided into
         a sub-chain of length $m$ and the ``environment''
         $R_{m,N}$.} 
\label{fig:1}
\end{figure}

Out of equilibrium, one can no longer settle with the reduced
basis set of ${\cal H}^{f}_{N}$ at temperature $T_N$, as the NEQ
dynamics involves all energy scales exceeding $T_N$. A complete
basis set of the Fock space ${\cal F}_N$ of ${\cal H}_N$ is
required. One possible choice of a complete basis set is given
by ${\cal F}_N = \{ |l,e;m \rangle_{tr} \}$, where $l$ labels
the NRG eigenstates of ${\cal H}^{f}_{m}$ {\em discarded} at
iteration $m = 1,\cdots, N$, and $e$ is a tuple of quantum
numbers labeling the ``environment'' degrees of freedom on
the remaining chain tail $R_{m,N}$, see Fig.~\ref{fig:1}.
For $m = N$, all NRG eigenstates are considered discarded.
Restricting attention to local operators $\hat O$ which do
not act on the ``environment'' degrees of freedom, the time
evolution of $\langle \hat O \rangle$ is formally given by
\begin{eqnarray}
\langle \hat O \rangle (t) &=& \sum_{m,m'} \sum_{l,l',e,e'}
        \, _{tr}\langle l,e;m| {\hat O} |l',e';m'\rangle_{tr}
        \times
\nonumber
\\
&& \ _{tr}\langle l',e';m'|
        { e^{-i{\cal H}^{f}_{N} t} \hat{\rho}_{eq}
                 e^{i{\cal H}^{f}_{N} t}}
          |l,e;m\rangle_{tr} \; .
\label{O-of-t}
\end{eqnarray}

The sum over $m$, $m'$ in Eq.~(\ref{O-of-t}) divides into
three parts: $m > m'$, $m = m'$, and $m < m'$. Consider
$m > m'$. Introducing the states $\{ |l, e; m\rangle_{kp} \}$,
where $l$ now labels the {\em retained} NRG eigenstates of
${\cal H}^{f}_{m}$ and $e$ records the ``environment''
degrees of freedom, one has the identity
\begin{equation}
\sum_{ {m', l', e'} \atop {(m'>m)}}
     |l', e'; m'\rangle_{tr}\ _{tr}\langle l', e'; m'| =
\sum_{l, e}
     |l, e; m\rangle_{kp}\ _{kp}\langle l, e; m| .
\label{identity}
\end{equation}
A similar identity applies to $m' > m$ upon interchanging the
roles of $m$ and $m'$. Inserting Eq.~(\ref{identity}) and its
$m \leftrightarrow m'$ counterpart into Eq.~(\ref{O-of-t})
one obtains
\begin{eqnarray}
\langle \hat{O} \rangle (t) &=&
        \sum_{m = 0}^{N}\sum_{r,s}^{trun} \;
        e^{i(E_{r}^m -E_{s}^m)t}
        O_{r,s}^m \rho^{red}_{s,r}(m) \; ,
\label{eqn:time-evolution}
\end{eqnarray}
where $O_{r,s}^m = \langle r,e;m| \hat{O}| s,e;m \rangle$
is independent of $e$, and $\rho^{red}_{s,r}$ is the
{\em reduced density matrix}~\cite{Feynman72,Hofstetter2000}
\begin{equation}
\rho^{red}_{s,r}(m) = \sum_{e}
          \langle s,e;m|\hat{\rho}_{eq} |r,e;m \rangle .
\label{eqn:reduced-dm-def}
\end{equation}
Here we have used the conventional NRG approximation 
${\cal H}^{f}_{N} |l,e;m \rangle \approx E_l^m |l,e;m \rangle$,
where $E_l^m$ is the NRG eigen-energy of ${\cal H}^{f}_{m}$
corresponding to the eigenstate $l$~\cite{Wilson75}.
The restricted sum
$\sum^{trun}_{r,s}$ in Eq.~(\ref{eqn:time-evolution})
implies that at least one of the states $r$ and $s$ is
discarded at iteration $m$.

Equation~(\ref{eqn:time-evolution}) is one of the central
results of this paper. In contrast to the
TD-DMRG~\cite{Marston2002,
DaleyKollathSchollwoeckVidal2004,WhiteEeiguin2004,
GobertTdDMRG2004} and a previous attempt at a time-dependent
NRG~\cite{Costi97}, {\em all states of the finite-size Fock
space are retained, and all energy scales are explicitly
taken into account. No basis set reduction is imposed.} 
To mimic the relaxation in an infinite-size system
we (i) average Eq.~(\ref{eqn:time-evolution}) over $N_z$
different realizations of the NRG discretization using
Oliveira's $z$-trick~\cite{YoshidaWithakerOliveira1990},
and (ii) introduce a scale-dependent damping factor,
\begin{equation}
e^{i(E_{r}^m - E_{s}^m) t} \to
         e^{i(E_{r}^m - E_{s}^m) t} e^{ -\alpha D_m t} ,
\end{equation}
for all $E_r^m \neq E_s^m$ in Eq.~(\ref{eqn:time-evolution}).
Here $\alpha$ is a constant of order unity, representing
a Lorentzian broadening of the NRG levels.
Below we present results with and without the additional
damping factor $\alpha$.

\paragraph{Algorithm.}
To implement the TD-NRG at temperature $T$, one first
selects a chain length $N$ such that $T \approx T_N$.
Two simultaneous NRG runs are then performed, one for
${\cal H}^i$ and another for ${\cal H}^f$, to obtain
(i) the equilibrium density matrix $\hat{\rho}_{eq}$,
(ii) the NRG eigen-energies and eigenstates of
${\cal H}^f$ and ${\cal H}^i$, and
(iii) the overlap matrices $S_{r,r'}(m)$ between the
eigenstates of ${\cal H}^i$ and ${\cal H}^f$ at each
iteration $m$. Using a recursion relation
$\rho^{red}(m) \to \rho^{red}(m-1)$, the reduced density
matrix of Eq.~(\ref{eqn:reduced-dm-def}) is evaluated
with respect to the NRG eigenstates of ${\cal H}^i$,
and then rotated to the NRG eigenstates of ${\cal H}^f$
using the overlap matrices $S_{r,r'}(m)$. The expectation
value $\langle \hat{O} \rangle (t)$ follows from
evaluation of Eq.~(\ref{eqn:time-evolution}).

\paragraph{Benchmark.}
As a critical test of our method, we first apply it to the
resonant-level model (RLM), describing the hybridization
of a localized level $d^{\dagger}$ with a band of spinless
conduction electrons $c^\dagger_k$:
\begin{equation}
{\cal H} = \sum_k \varepsilon_k c^\dagger_k c_k
           + E_d(t) d^\dagger d 
           + V \sum_k
               \{d^\dagger c_k + c^\dagger_k d \} .
\label{eqn:res-level-model}
\end{equation}
Choosing $\hat{n}_d = d^\dagger d$ as the observable
$\hat{O}$, we consider a stepwise change in the energy of
the level: $E_d(t) = \theta(-t) E_d^0 + \theta(t) E_d^1$.
In the wide-band limit,
$n_d(t) = \langle \hat{n}_d \rangle (t)$ can be solved
exactly in closed analytical form using the Keldysh formalism.
For $T = 0$, the solution features an exponential decay from
the initial equilibrium occupancy of ${\cal H}^i$ to the new
equilibrium occupancy of ${\cal H}^f$ with two decay rates
$\Gamma$ and $2\Gamma$. Terms decaying at rate $\Gamma$
show an additional modulation of frequency
$|E_d^1|$~\cite{exact_solution}. 
Here $\Gamma = \pi \rho_0 V^2$ is the hybridization width,
$\rho_0$ being the conduction-electron density of states at
the Fermi energy. 

\begin{figure}[tb]
\centering
\includegraphics[height=60mm]{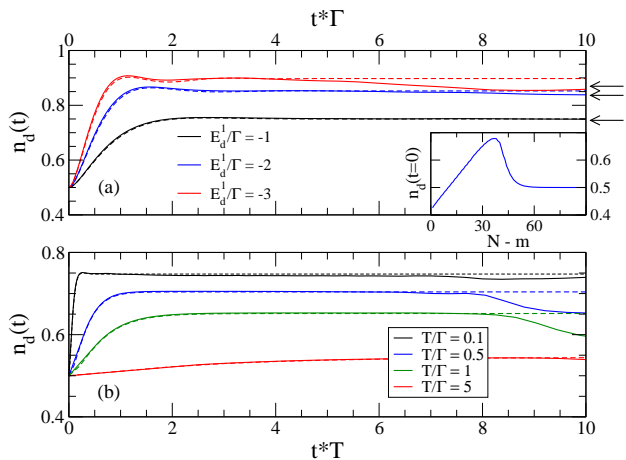}
\caption{Time-dependent occupancy of the RLM for
      (a) $E_d^0 = 0$, $T \to 0$ and three different
      values of $E_d^1/\Gamma = -1, -2, -3$, and (b)
      $E_d^0 = 0$, $E_d^1 = -\Gamma$ and four different
      temperatures $T/\Gamma = 0.1, 0.5, 1, 5$.
      Solid lines are the results of the TD-NRG with
      $\Lambda = 1.3$, $D/\Gamma = 500$, $N_s = 1000$,
      $N_z = 32$, and $\alpha = 0$ (no damping factor).
      Dashed lines are the exact solution in the wide-band
      limit. Right arrows mark the saturated NRG occupancies for
      $\alpha > 0$. Inset to (a): Progressive calculation
      of $n_d(0)$ as a function of the number of backward
      iterations $N-m$, for $E_d^1 = -2\Gamma$.}
\label{fig:2}
\end{figure}

Figure~\ref{fig:2} shows a comparison of the TD-NRG with
the exact analytical solution for $n_d(t)$, in response to
sudden change in the energy of the level from $E_d^0 = 0$
to $E_d^1 \neq 0$. Here we have used a symmetric box
density of states $\rho(\varepsilon) = \rho_0
\theta (D - |\varepsilon|)$ for the TD-NRG~\cite{Wilson75},
and averaged over $N_z = 32$ equally spaced values of
$z$~\cite{YoshidaWithakerOliveira1990}. No damping factor
was introduced. To facilitate comparison with
the wide-band limit, a large bandwidth $D/\Gamma = 500$
was used.

As seen in Fig.~\ref{fig:2}(a), the TD-NRG accurately
reproduces the exact $T = 0$ solution {\em on all time scales}
up to moderately large $E_d^1$. For $E_d^1 = -\Gamma$, the
maximal error is less than 1\%. For $E_d^1 = -2\Gamma$ it
is less than 2.5\%.
The short-time behavior is well described by the TD-NRG
for all $E_d^1$. In particular, $n_d(t \to 0^+)$ coincides
with the initial equilibrium NRG occupancy of ${\cal H}^i$
independent of $E_d^0$ and $E_d^1$. As demonstrated in the
inset to Fig.~\ref{fig:2}(a), summation over the states of
all NRG iterations is essential for recovering $n_d(0)$.
One cannot settle with just a  single NRG iteration as in
the equilibrium case.

Remarkably, the TD-NRG retains the same accuracy in
Fig.~\ref{fig:2}(a) up to arbitrarily long times. Deviations
from the new equilibrium occupancy of ${\cal H}^f$ are
only slight at large $t$, and are further reduced by
decreasing $\Lambda$, due to elongation of the effective
chain length corresponding to the energy scale $E_d^1$.
The long-time deviations in $n_d(t)$ stem from a difference
between the time-evolved density matrix in a finite-size
system and the equilibrium density matrix
$\hat{\rho} \propto e^{-\beta {\cal H}^f}$, approached
only for $\Lambda\to 1^+$.

The TD-NRG can also be applied to finite temperatures,
see Fig.~\ref{fig:2}(b). For moderate $E_d^1$, there is
excellent agreement with the exact curves up to
$t$ several times larger than $1/T$, at which point
finite-size oscillations develop. The latter oscillations
are greatly reduced by averaging over the different
$z$'s, to the extent that they appear damped
at longer times. Note, however, that all curves in
Fig.~\ref{fig:2}(b) have practically saturated at their
new equilibrium values prior to the appearance of
oscillations.

We stress that no explicit damping factor was utilized
in Fig.~\ref{fig:2}. A nonzero $\alpha$ depresses the
oscillations in $n_d(t)$, which saturates at the new
steady-state values indicated by arrows in Fig.~\ref{fig:2}(a).
The difference $\delta n_d$ between the new equilibrium
occupancy of ${\cal H}^f$ and the steady-state value reached
serves as a sharp criterion for the accuracy of the TD-NRG.
The smaller $\delta n_d$ is the better the TD-NRG performs
on all time scales. This provides one with a valuable
estimate for the accuracy of the TD-NRG in cases where
no other reference exists.

\paragraph{Single impurity Anderson model.}

The first nontrivial model which lacks an exact solution
out of equilibrium is the single impurity Anderson model
(SIAM), describing a spinful resonant level with an on-site
repulsion $U$:
\begin{eqnarray}
{\cal H} &=& \sum_{k, \sigma} \varepsilon_{k\sigma}
                   c^\dagger_{k\sigma} c_{k\sigma} 
             + \sum_{\sigma = \pm 1}
                   \left[
                        E_d(t) - \frac{\sigma}{2} H(t)
                   \right ]
                   d^\dagger_\sigma d_\sigma 
\nonumber \\
&& + U \hat{n}^d_\uparrow \hat{n}^d_\downarrow
            + V(t) \sum_{k, \sigma}
              \left\{
                   d^\dagger_\sigma c_{k\sigma} +
                   c^\dagger_{k\sigma} d_\sigma
              \right \}
\label{eqn:siam}
\end{eqnarray}
with $\hat{n}_{\sigma}^{d} = d^{\dagger}_{\sigma} d_{\sigma}$.
This Hamiltonian is commonly used to model a single
Coulomb-blockade resonance in ultra-small quantum
dots. To probe the relaxation both in the spin and charge
sectors, we consider in the following a simultaneous
stepwise change in the energy of the level,
$E_d(t) = \theta(-t) E_d^0 + \theta(t) E_d^1$, and in
the local magnetic field,
$H(t) = \theta(-t) H_0 + \theta(t) H_1$. We also
permit a sudden change in the hybridization width
$\Gamma(t) = \pi \rho_0 V^2(t)$:
$\Gamma(t) = \theta(-t) \Gamma_0 + \theta(t) \Gamma_1$.
As our prime interest is in the relaxation to a new
Kondo state, we focus hereafter on $H_1 = 0$ and only
briefly comment on nonzero $H_1$.

\begin{figure}[bt]
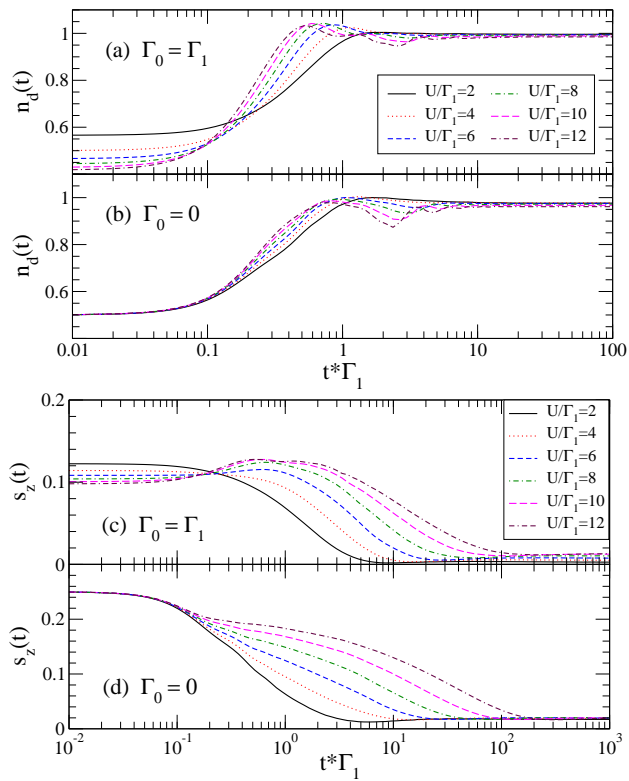

\centering
\includegraphics[height=50.8mm,clip=true]{fig3ab}

\includegraphics[height=51.6mm,clip=true]{fig3cd}
\vspace{-5pt}
\caption{Time-dependent occupancy $n_d(t)$, Figs.~(a) and
         (b), and magnetization $S_z(t)$, Figs.~(c) and
         (d), of the SIAM with $T \to 0$,
         $D/\Gamma_1 = 20$, $H_0 = 2E_d^0 = \Gamma_1$,
         $H_1 = 0$, $E_d^1 = -U/2$, and six different
         values of $U/\Gamma_1 = 2, 4, 6, 8, 10, 12$.
         Here $\Gamma_0 = \Gamma_1$ in Figs.~(a) and (c),
         whereas $\Gamma_0 = 0$ in Figs.~(b) and (d). NRG
         parameters: $\Lambda = 1.8$, $N_s = 1000$,
         $N_z = 16$, and $\alpha = 0.1$.}
\label{fig:3}
\end{figure}

In the spirit of the experiment of Elzerman
{\em et al.}~\cite{Kouwenhoven2004}, we begin with a
spin-polarized impurity, where the lower spin state
is degenerate with the empty-orbital state:
$E_d^0 - H_0/2 = 0$. At $t=0$, both the local magnetic
field is switched off and the energy of the level is
shifted to $E_d^1 = -U/2$. The resulting time evolutions
of the impurity occupancy
$n_d(t) = \langle \hat{n}^d_{\uparrow} +
\hat{n}^d_{\downarrow} \rangle (t)$ and magnetization
$S_z(t) = \frac{1}{2} \langle \hat{n}^d_{\uparrow} -
\hat{n}^d_{\downarrow}\rangle (t)$ are plotted in
Fig.~\ref{fig:3}, for $T \to 0$, $H_0 = \Gamma_1$,
and different values of $U$. Two separate initial
conditions are considered: $\Gamma_0 = \Gamma_1$ and
$\Gamma_0 = 0$, representing different relaxations
to the strong-coupling fixed point. The former
corresponds to relaxation from a spin-polarized
mixed-valent state, while the latter describes
evolution from a free-impurity fixed point.

Two distinct time scales are clearly visible in
Fig.~\ref{fig:3}, for the relaxation of spin and charge.
Similar to the RLM, $n_d(t)$ equilibrates on a time
scale $t_{\rm ch} \propto 1/\Gamma_1$, and develops
Rabi-type oscillations for $|E_d^1| > \Gamma_1$. For
$t \gg t_{\rm ch}$, the occupancy must
saturate at its new equilibrium value $n_d = 1$. This
behavior is well captured by the TD-NRG, attesting to
its accuracy. The long-time deviations $\delta n_d$
are less than $0.015$ for $\Gamma_0 = \Gamma_1$ and less
than $0.04$ for $\Gamma_0 = 0$. Importantly, $\delta n_d$
is notably smaller for the SIAM than for the RLM with
comparable $E_d^1$.

\begin{figure}[tb]
\centering
\includegraphics[height=52.2mm,clip=true]{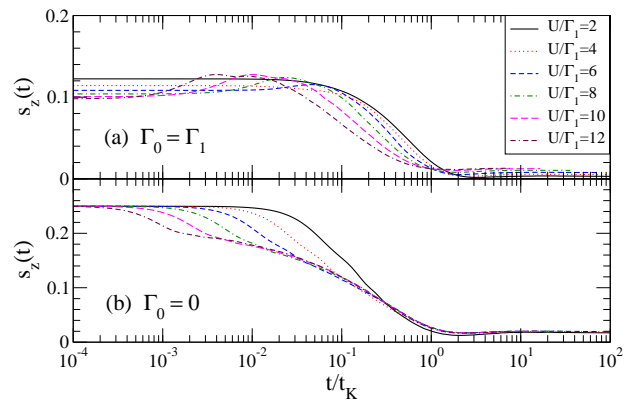}
\vspace{-5pt}
\caption{The impurity magnetization curves of
         Fig.~\ref{fig:3}, replotted versus $t/t_K$
         with $t_K = 1/T_K$. Here $\Gamma_0 = \Gamma_1$
         in Fig.~(a), while $\Gamma_0 = 0$ in Fig.~(b).}
\label{fig:4}
\end{figure}

Contrary to the dynamics of $n_d(t)$, the spin magnetization
$S_z(t)$ equilibrates on a much longer time scale
$t_{\rm sp}$. To elucidate the strong $U$ dependence of
$t_{\rm sp}$, we replotted the magnetization curves of
Fig.~\ref{fig:3} versus $t/t_K$ with $t_K = 1/T_K$, see
Fig.~\ref{fig:4}. Here $T_K$ is the Kondo temperature
of ${\cal H}^f$, defined by
$T_K\chi(T_K) = 0.07$~\cite{KrishWilWilson80a} ($\chi$
being the impurity susceptibility of ${\cal H}^f$).
While the initial dynamics of $S_z(t)$ is governed
by $t_{\rm ch} \propto 1/\Gamma_1$, the eventual
relaxation is clearly governed by the Kondo time scale:
$t_{\rm sp} \propto 1/T_K$~\cite{LobaskinKehrein2004}.
This is best seen for $\Gamma_0 = 0$,
Fig.~\ref{fig:4}(b), where all curves gradually
collapse onto a single master curve, provided a
stable local moment was formed on the level. For
$\Gamma_0 = \Gamma_1$, there is no universality
in $t/t_K$. In fact, $S(t)$ differs in form from
that of $\Gamma_0 = 0$, revealing a sensitivity
of spin relaxation to the initial conditions imposed
on the system. A nonzero $H_1$ suppresses the Kondo
effect. Accordingly, the spin relaxation time
$t_{\rm sp}$ is reduced, approaching $t_{\rm ch}$
for $H_1 \gg T_K$ (not shown).

\paragraph{Summary and discussion.}

A powerful new approach to the nonequilibrium dynamics
of QIS was presented based upon Wilson's NRG. All states
of the Wilson chain are summed over in this scheme, and no
basis set reduction is imposed. We established the accuracy
of the approach on all time scales by comparison to the
exact solution of the RLM, and demonstrated its ability to
access exceedingly long time scales by investigating the
equilibration of the SIAM. The spin relaxation time
$t_{\rm sp}$ of the SIAM was shown to be governed by the
Kondo time scale $t_K = 1/T_K$, for $T \ll T_K$. However,
there is no universality in $t/t_K$. Spin relaxation is
sensitive to the initial conditions imposed on the system.
Although our specific examples were focused on Fermionic
baths, the TD-NRG can equally be applied to Bosonic
baths~\cite{BullaBoson2003}. This should allow for an
accurate investigation of the time evolution of a
two-level system~\cite{Leggett1987}, as relevant to
quantum computers, and of biological donor-acceptor
molecules. Finally, we expect that our approach can
be generalized to continuous time dependences (such
as ac drives) by discretizing the time axis, and
evaluating Eq.~(\ref{eqn:time-evolution}) consecutively
using the discrete set of Hamiltonians ${\cal H}^f(t_j)$.
This will provide a very general approach to the
real-time dynamics of
quantum-impurity systems at any temperature $T$.

We have benefited from discussions with R.~Bulla,
G.~Czycholl, S.~Kehrein, K.~Ingersent, D.~Vollhardt, M.~Vojta and
especially E.~Lebanon. F.B.A. acknowledges funding of
the NIC, Forschungszentrum J\"ulich, under project
no. HHB000. A.S. was supported in part by the Centers
of Excellence Program of the Israel Science Foundation.

\end{document}